\documentclass[onecolumn]{aa} 
\usepackage{graphicx}
\usepackage{txfonts}
\usepackage{color}

\usepackage{natbib}
\bibpunct{(}{)}{;}{a}{}{,} 
\usepackage{graphicx}
\usepackage{lscape}
\usepackage{rotating}

\def\simle{\lower 2pt \hbox {$\buildrel < \over {\scriptstyle \sim }$}}
\def\simge{\lower 2pt \hbox {$\buildrel > \over {\scriptstyle \sim }$}}

\begin{document}

\title{The catalog of nearby black hole candidates}

\author{Lauren\c{t}iu I. Caramete
	\inst{1,2,}\thanks{Member of the International Max Planck Research School
(IMPRS) for Astronomy and Astrophysics at the Universities
of Bonn and Cologne}
		\and Peter L. Biermann\inst{1,} \inst{3,} \inst{4,} \inst{5,} \inst{6}
		}
\institute{Max Planck Institute for Radio Astronomy, Auf dem H\"ugel 69, 53121 Bonn, Germany
	\and
Institute for Space Sciences, P.O.Box MG-23, Ro 077125 Bucharest-Magurele, Romania
	\and
Department of Physics and Astronomy, University of Bonn, Endenicher Allee 11-13, Bonn, Germany
\and
Department of Physics and Astronomy, University of Alabama, Tuscaloosa, Alabama 35487, USA
\and
Department of Physics, University of Alabama at Huntsville, 301 Sparkman Drive Huntsville AL 35899, USA
\and
Karlsruhe Institute of Technology, P.O. Box 3640, 76021 Karlsruhe, Germany}

%

\abstract{In order to study the association of the origin of ultra high
 energy cosmic rays (UHECR) with active galactic nuclei (AGN) at all levels of their
activity we require an unbiased sample of black holes.}{Here we describe
such a sample, of about 6 000 black holes, within the local Universe, inside the GZK (Greisen Zatsepin Kuzmin) limit, around 100 Mpc.}{
 The starting point is the 2 micron all sky survey, with the next steps as:
 test its completeness down to low flux densities, confine it to redshifts
 $z < 0.025$, limit it to early Hubble type galaxies, test
 with B-V colors and with the FIR/Radio ratio the possible separation in classes
 of sources for the ultra high energy cosmic rays, use the spheroidal stellar
 component - black hole mass relationship to derive black hole masses, and test them with known
 black hole masses.}{The statistics are consistent with the mass function of
black hole masses, with a relatively flat distribution to about $10^{8} \, M_{\odot}$, and thereafter a very steep spectrum. Our sample cuts off
 just below $10^{7} \, M_{\odot}$, indicating a gap for lower masses,
 below somewhere near $10^{6} \, M_{\odot}$. We construct the sky
distribution for all black holes above $10^{7} \, M_{\odot}$, $10^{8} \, M_{\odot}$, and $3\times 10^{8} \, M_{\odot}$,
and also show the sky distribution for the five
redshift ranges 0.0 - 0.005, 0.005 - 0.0010, 0.010 - 0.015, 0.015 -
0.020, and 0.020 to 0.025.}{}


\authorrunning{L.I. Caramete and P.L. Biermann}
\titlerunning{The catalog of nearby black hole candidates}
   \maketitle

\section{Introduction}

The samples of active galactic nuclei are small in the nearby universe and incompletely distributed on the sky, for example, there are only 865 objects in the V{\'e}ron catalog inside a 100 Mpc region around our Galaxy \citep{2010A&A...518A..10V}.
 As many models developed for the determination of the origin of UHECR require information concerning the past activity of a black hole, and little of the current activity, (see, e.g.  \citet{1999MNRAS.307..491B,2000MNRAS.316L..29B,2002APS..APRB17041B,2002APh....16..265L,2002PhRvD..66g4014M,2008MNRAS.383..663B,1977MNRAS.179..433B,2007APh....27..473L}) it is useful to derive a nearly complete, all sky sample of black holes in the nearby Universe, inside the sphere around our Galaxy determined by the GZK distance, \citet{1966PhRvL..16..748G,1966ZhPmR...4..114Z}.

We note that in fact all black holes can be detected as active, some at very low level, most easily via non-thermal radio emission \citep{1984A&A...130L..13P,1984ApJ...283..479E,2000ApJ...542..186N}.
One very good example is the black hole at the center of our Galaxy, it’s past activity was discovered by looking at the entire electromagnetic emission \citep{2010PNAS..107.7196M}.

The overall strategy of finding the associations between the astrophysical
 sources and the events observed in the sky is to develop a nearly complete sample
of possible sources, separate them in sub-samples according to their nature (e.g. blue or red
 star dominated, starburst or non-starburst galaxies), take the maximal sources from each
 set and model the physical processes that are involved in the production of the energetic
 particles in order to obtain the maximal flux of the cosmic rays. Then the next step is to model and implement
 the propagation of these particles from the distribution of sources to their arrival
 distribution here on Earth including the deflection in a structure of magnetic fields. Finally one needs to make the statistical correlation between the virtual events and the real observed ones.
   In this article the first steps are taken with the next ones treated in the following articles.

\section{The active black holes}

In a recent article, \citet{2010A&A...521A..55C}, we presented the most important results concerning the catalog of black hole candidates which include the mass function of the black holes and its consequences, together with the method of determination. Here we wish to present the catalog with all its fields and informations and the associated analysis and plots concerning the characteristics of the catalog which we were unable to include in the previous paper.

Starting with the Two Micron All Sky Survey (2MASS), \citet{2006AJ....131.1163S} in the near-infrared wavelength domain from NED (NASA/IPAC Extragalactic Database) we build a catalog of massive black holes.

\begin{figure}[htpb]
\centering
\includegraphics[viewport=0cm 0cm 20cm 10cm,clip,scale=0.8]{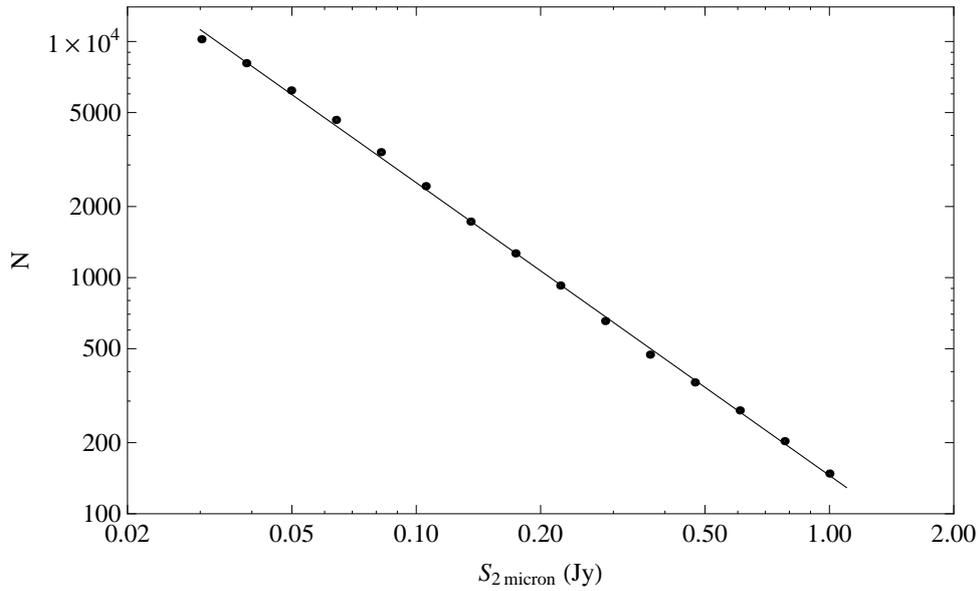}
\caption{Integral source counts, for 10 284 NED candidate sources in
 the case of selection at 2 micron, $z \leq 0.025$. There is no break in the distribution indicating incompleteness.}
\label{2microni10284obj}
\end{figure}

We start with a complete sample, down to 0.03 Jy, as implied by the
$\log N - \log S$ distribution, Fig.~\ref{2microni10284obj}, cumulative number $N$ down to a sufficient low flux density $S$ to give us a big sample and still be complete. This gives 10 284 sources.

\begin{figure}[htpb]
\centering
\includegraphics[viewport=3cm 0cm 20cm 10cm,clip,scale=0.8]{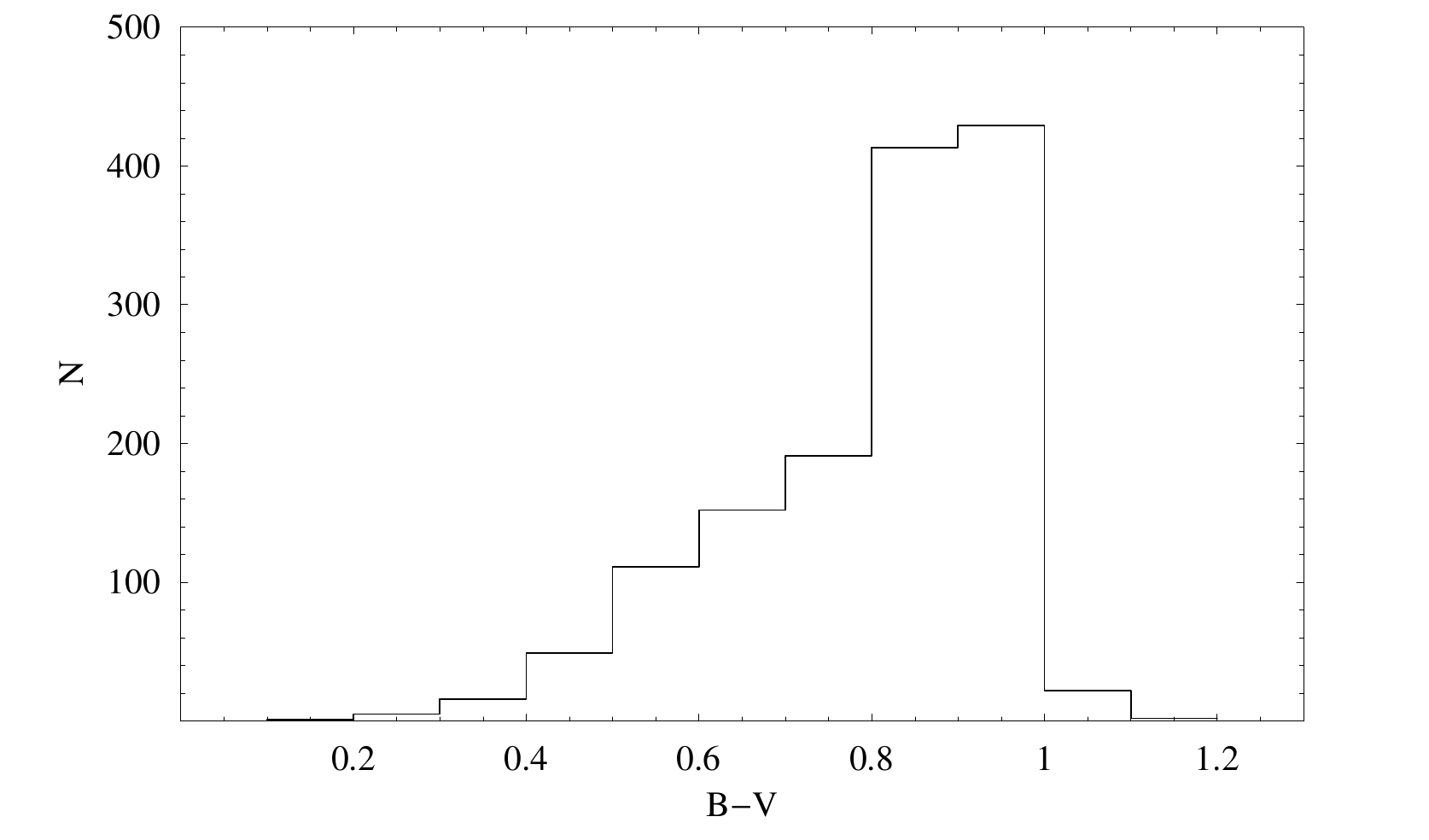}
\caption{Histogram for available B-V from the full sample of NED candidate sources
 at 2 micron. From 5 894 galaxies in the full sample only around $25\% $ have color index.}
\label{HistogramBVBH}
\end{figure}

As a first step we try to separate the galaxies by activity (which contain a starburst or not), or stellar type (if the stellar type is dominated by old red stars or blue young ones). This failed because to lack of informations regarding the FIR/Radio ratio and B-V color index, too few observations in radio and infrared or visible. Then we notice that the only information given about these galaxies for all sources is
the Hubble-type. Because of this we make a such selection as to include all early Hubble-type galaxies including the Hubble-type Sb.
 We included cD galaxies, but we rejected all Sbc and later Hubble-type galaxies like SAc for example, and all known starburst
galaxies. In this rather conservative selection we exclude several
 known Seyfert galaxies. The final list has 5 894 galaxies and we name this the catalog of nearby black hole candidates(BH-Cat).
As a control we repeat the B-V histogram of the few for which this
information is known in Fig.~\ref{HistogramBVBH}.

This demonstrates that there is a tail toward bluer galaxies,
presumably those in a merger; however, most galaxies, for which B-V is
known, are quite red, as expected for early Hubble type galaxies.
Unfortunately, B-V is known only for a relatively small subset of all
the galaxies, for only 1488, and of these 980 have B-V bigger than 0.8.

Using the black hole mass with spheroidal stellar population
relationship \citet{1997AJ....114.1771F,1998A&A...334...87W,1998A&A...331L...1S,1998AJ....115.2285M}, the estimation of
 their black hole mass can be readily made like:

\
 \begin{equation}
  M_{BH} \propto f_{2\mu} D^{2}
  \label{eqno1}
 \end{equation}
\

Since the estimate uses the redshift, which is a bad indicator
 of true distance at distances less than that of the Virgo cluster, we use the available distances
 provided by the work of Barry F. Madore and Ian P. Steer (available at {\it http://nedwww.ipac.caltech.edu/level5/NED1D/intro.html}) who compiled a database of 3 065 accurate, contemporary distances to 1 073 galaxies with modest recessional velocities (that is, less than 1/8 c) published between almost exclusively 1990 and 2006.
 From this we match 429 distances in the catalog of massive black hole, and for the rest of distances less than that of the Virgo cluster we used distances with respect to the Virgo infall only provided by NED.

We construct two independent lists of black hole masses from the literature, one which is used to derive the proportionality coefficients (from \citet{2004ApJ...604L..89H,2003ApJ...589L..21M,2001ASPC..249..335M}) and one to check the known mass from the literature with the estimated one derived in this catalog (from \citet{2006AJ....131.1236D,2006ApJ...641L..21G,2010MNRAS.407.2399M}). First we divide the BH-Cat in five main categories, called E, S0, Sa, Sb, Sab and we use the first list from the literature to calculate the proportionality coefficients for each of the five categories. Then we use equation~\ref{eqno1} together with this coefficients to calculate the black hole mases and their errors. Just from Poisson statistics the error bar in these determinations is already a few percent,
and from systematics such as black holes in later Hubble type galaxies, and mis-estimated and therefore mis-allocated masses (i.e. inserted into the wrong bin) the error bar could be as high as $50 \%$.

\begin{figure}[htpb]
\centering
\includegraphics[viewport=0cm 0cm 20cm 11cm,clip,scale=0.9]{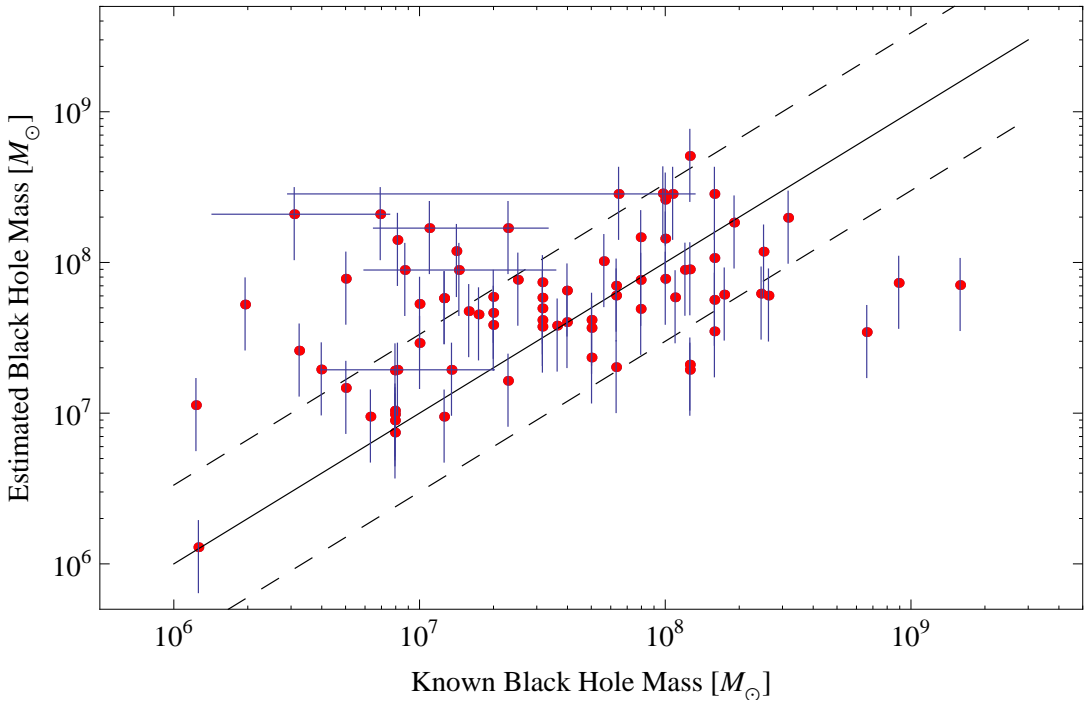}
\caption{Comparison between the masses of black holes from the literature and the black hole mass estimated using our approach, these masses have been systematically corrected for Hubble-type.}
\label{KnownOverEstimatedBHMas}
\end{figure}

We use the second list to check our estimates, in Fig.~\ref{KnownOverEstimatedBHMas}, showing the match between the literature mass
black hole and the estimated one using our formula. We have here with the solid line the simple linear function for which the two masses coincide, the 1$:$1 relation and with the dash lines the parallels indicating a mismatch of 0.3 above and below, which means not bigger than the factor of order 2.

\begin{figure}[htpb]
\centering
\includegraphics[viewport=0cm 0cm 29cm 11cm,clip,scale=0.9]{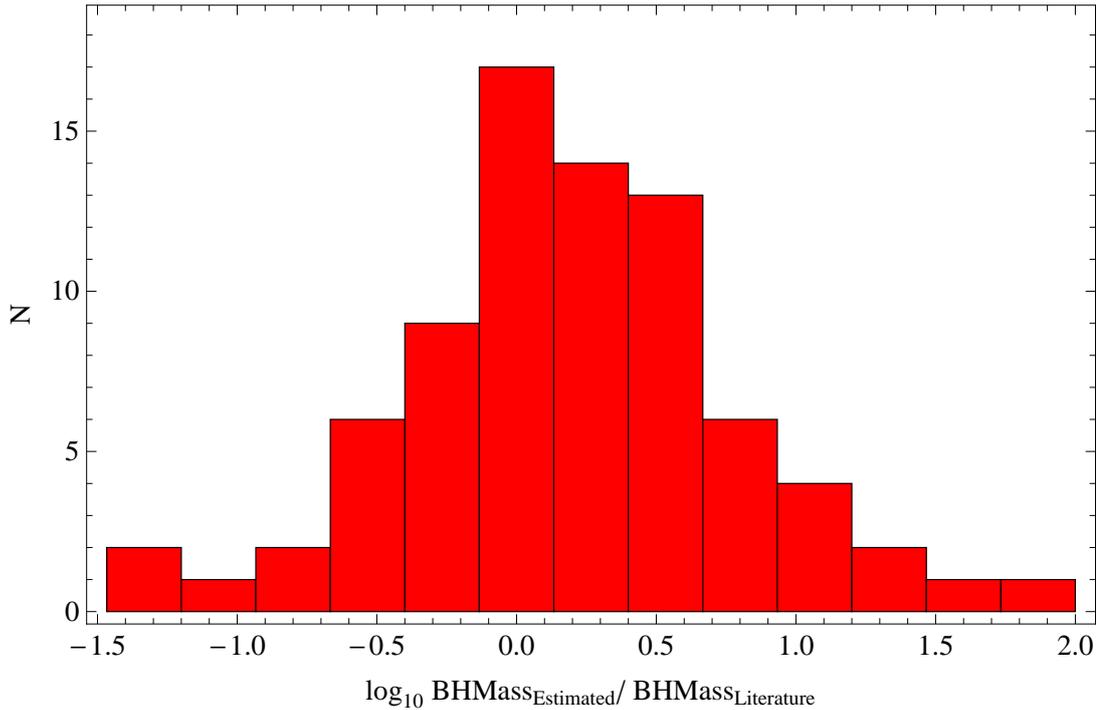}
\caption{Histogram of the ratio between the estimated black hole mass and the litterature one using the second list. Here the mean of the distribution is 0.19 with a standard deviation of 0.6}
\label{HistogramEstimBHLittBH}
\end{figure}

This comparison illustrates that the formula of calculating the black hole mass works quite well, Fig.~\ref{KnownOverEstimatedBHMas}, Fig.~\ref{HistogramEstimBHLittBH}, specially in the region between $10^{7}\, M_{\odot}$ and $10^{8}\, M_{\odot}$, where most of black hole in the catalog are, making
 the use of the catalog of the massive black hole possible as a statistical tool in tracing the evolution and
current status of the structures at large scale. This can be readily used in cosmic ray studies with the observation that future
work will improve and tune the catalog to provide more accurate data.

For spiral galaxies and lenticular ones, the simple formula~\ref{eqno1}, gives a slightly higher mass suggesting that the 2 micron flux has contributions in this case from younger stellar
populations; the suggested mismatch corresponds to an average factor of about 2 - 3. Here with spiral galaxies we refer to all the SAa, SAb, SABa SABb and also SBa, SBb so including the galaxies with bar-like structure, extending from the central bulge and with lenticular galaxies the ones which have S0 as morphological type.

This list of 5 894 sources thus represents a cautious list of massive
black holes in the local universe; as we have eliminated a small number
 of known AGN with later Hubble types, the list is not complete for all
black holes. It is, however, an unbiased list, with whatever selection
effects there are in the 2MASS survey, constant across the entire
sky with the exception of course, of the Galactic Plane. As the formula for calculation the masses of the black holes is quite general, the disadvantage
 given by the poor precision is compensated by the large sample of the black hole list. This
 will give the possibility to make statistical analysis about the distribution of the
 black hole in the nearby universe.
 The list could be very much refined with further work, which would
 require new observations, best perhaps radio interferometric
 observations.

The data also show that we have very few black holes below $10^{7} \, M_{\odot}$. Cutting at $10^{7} \, M_{\odot}$ decreases the sample
 to only 5 619 from 5 894 black holes or galaxies.

As an example we present in table~\ref{table10SMBH} a sample of twenty galaxies as an illustration of the data available at the CDS, and contains the following information: Column 1 lists the name of the source, Column 2 gives the galactic longitude, Column 3 gives the galactic latitude, Column 4 gives the redshift, Column 5 provides the Hubble type of the galaxies, Column 6 gives the distance in Mpc, Column 7 gives the estimated mass of the black hole and it's error, Column 8 gives the B$-$V color index and the final column, Column 9 gives the FIR/Radio ratio (FIR = flux density at 60$\mu$ and Radio = flux density at 5GHz), together with errors or limits when available, this last one being a very good indicator of starburst activity. Unfortunately we have only limited data for the flux density, future observations will make possible further classification and refinement.

\section{Redshift distribution}

The following step is to subdivide the redshift range into bins of 0.05
 each, and show first the sky distribution for each bin separately, and
then in combination, with one color for each redshift bin. The color
sequence is Black, Blue, Green, Orange, and Red, for the intervals in
redshift 0.000 - 0.005, 0.005 - 0.010, 0.010 - 0.015, 0.015 - 0.020,
 and 0.020 to 0.025, respectively.

\begin{figure}[htpb]
\centering
\includegraphics[viewport=0cm 0cm 29cm 11cm,clip,scale=0.8]{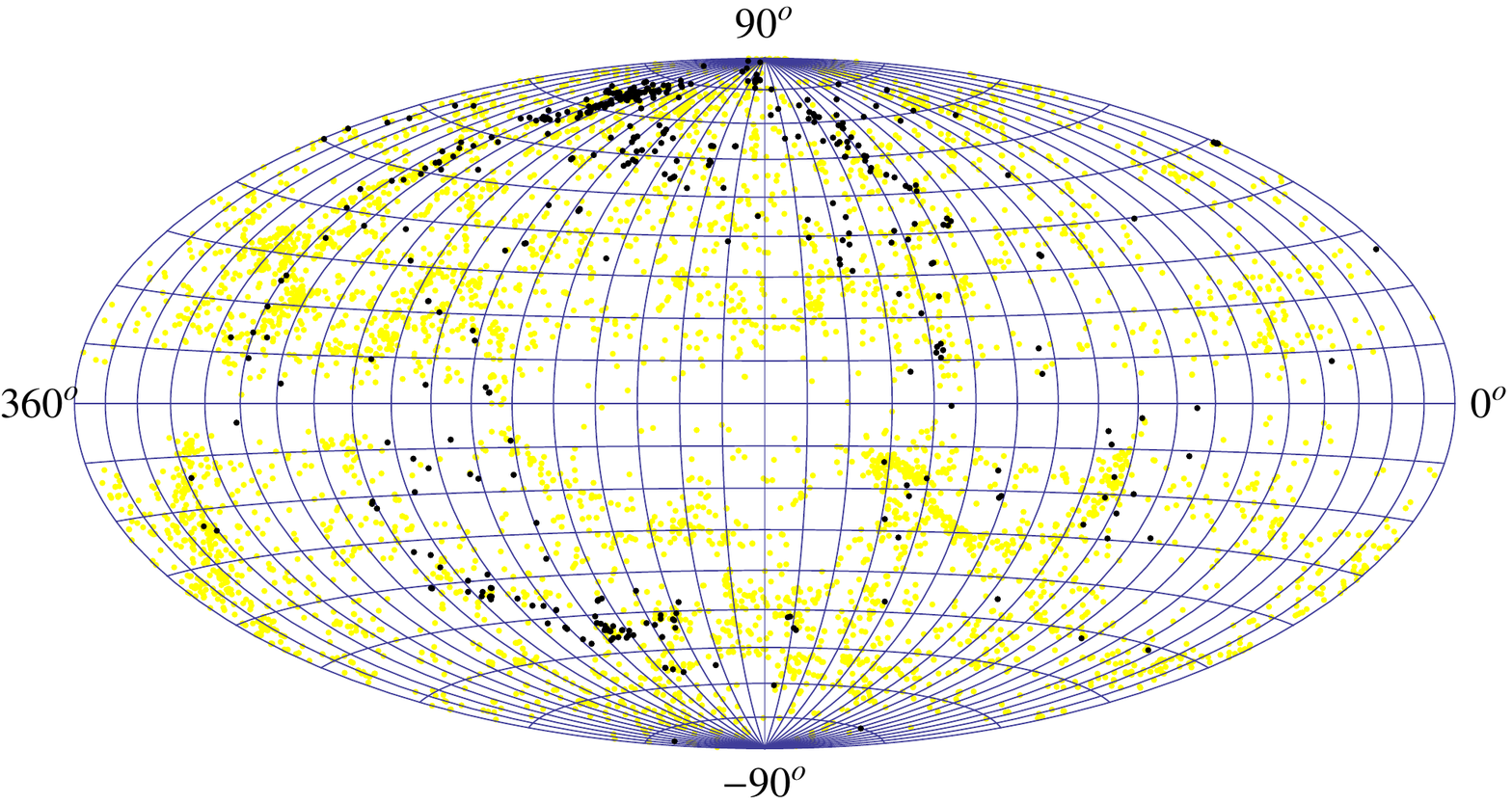}
\caption{Aitoff projection in galactic coordinates of 5 894 NED
 candidate sources. The color code is Black, corresponding to
redshifts between 0 - 0.005, and Yellow between 0.005 - 0.025. The choice of color is made to enhance the nearby population of massive black holes}
\label{SkyPlot2micronBLSelectionColorCodedBlack}
\end{figure}

\begin{figure}[htpb]
\centering
\includegraphics[viewport=0cm 0cm 29cm 11cm,clip,scale=0.8]{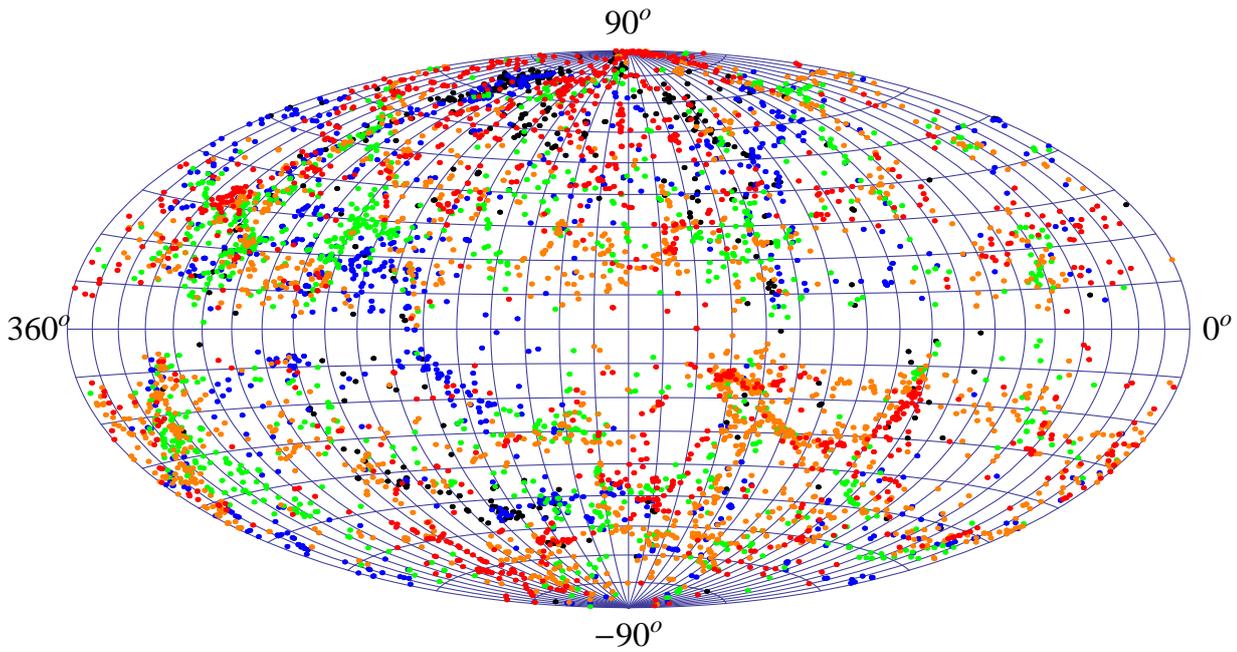}
\caption{Aitoff projection in galactic coordinates of 5 894 NED
 candidate sources. The color code is Black, Blue, Green,
 Orange, Red corresponding to redshifts between 0, 0.005, 0.01, 0.015, 0.02, 0.025.}
\label{SkyPlot2micronBLSelectionColorCoded}
\end{figure}

The first plot, Fig.~\ref{SkyPlot2micronBLSelectionColorCodedBlack}, illustrates the structure of the nearby population of massive black holes, with the supergalactic plane clearly visible as we trace the filaments represented with black color from left to the right.
The following sky plot, Fig.~\ref{SkyPlot2micronBLSelectionColorCoded}, is presenting in a simple colorful way the structure of the massive black holes as we go further and further in the redshift, complex filamentary structure appear indicating the non-uniformity of the massive black hole distribution in space.

\section{Selection by mass}

\begin{figure}[htpb]
\centering
\includegraphics[viewport=0cm 0cm 29cm 11cm,clip,scale=0.8]{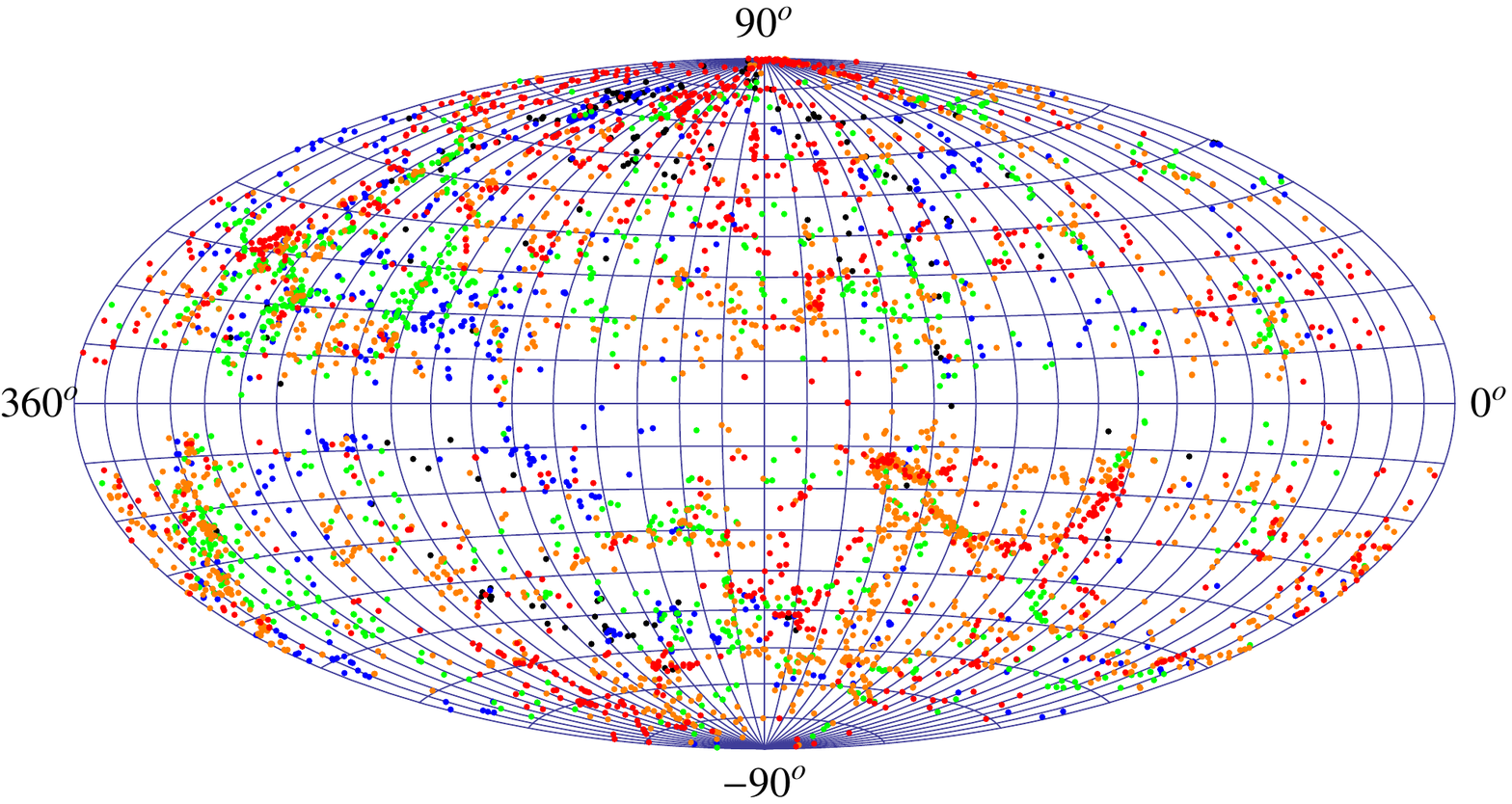}
\caption{Aitoff projection in galactic coordinates of a sample of 4832 objects from the black hole catalog with masses bigger than $3\times 10^{7}\, M_{\odot}$ . The color code is Black, Blue, Green,
 Orange, Red corresponding to redshifts between 0, 0.005, 0.01, 0.015, 0.02, 0.025. }
\label{SkyPlot2micronBLSelectionColorCodedForMassBiggerThan310To7}
\end{figure}

\begin{figure}[htpb]
\centering
\includegraphics[viewport=0cm 0cm 29cm 11cm,clip,scale=0.8]{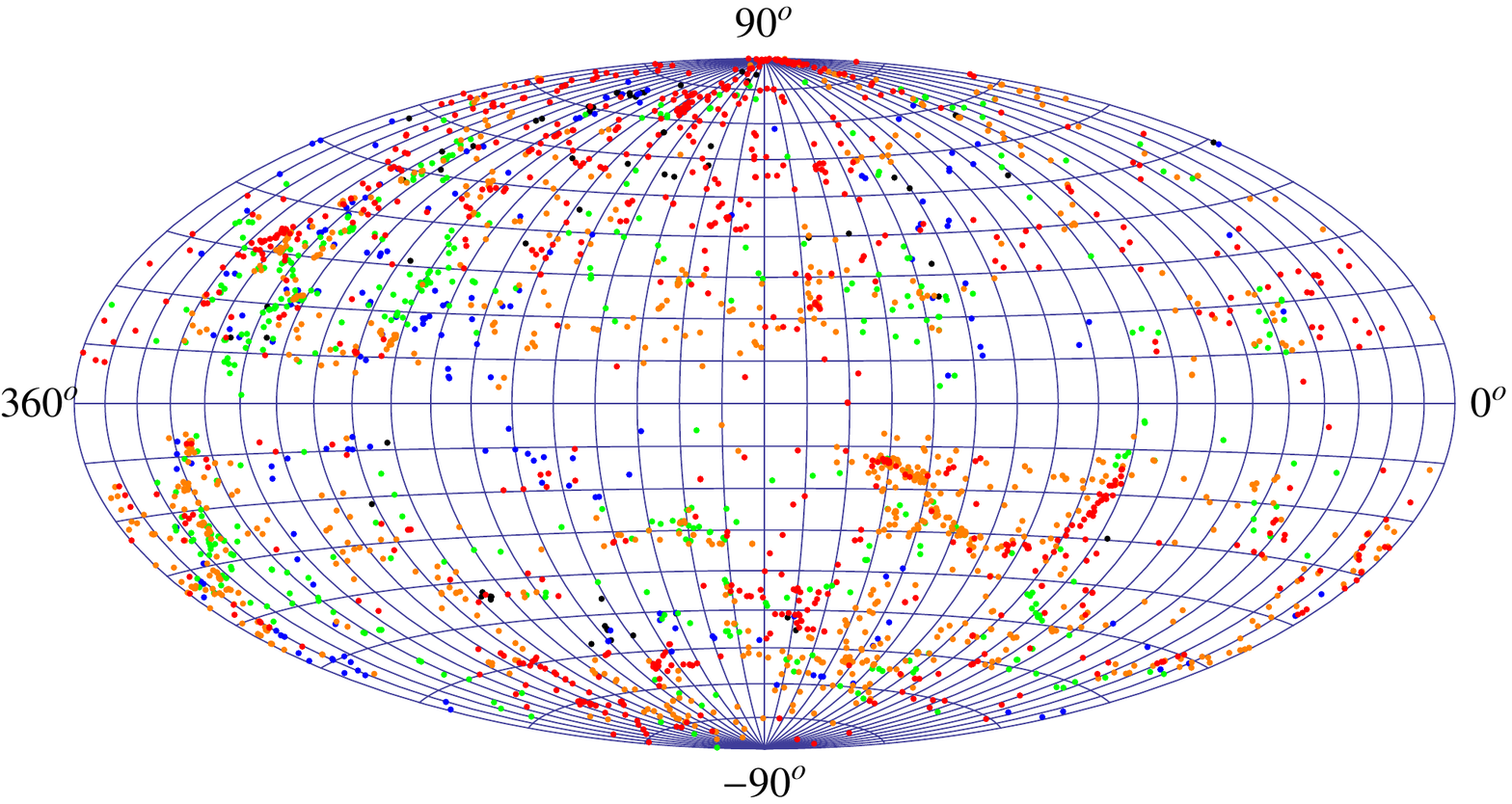}
\caption{Aitoff projection in galactic coordinates of a sample of 2546 objects from the black hole catalog with masses bigger than $10^{8}\, M_{\odot}$ . The color code is Black, Blue, Green,
 Orange, Red corresponding to redshifts between 0, 0.005, 0.01, 0.015, 0.02, 0.025. }
\label{SkyPlot2micronBLSelectionColorCodedForMassBiggerThan10To8}
\end{figure}

This section presents the overall characteristics of several cuts in the initial massive black hole list. We show in the next plots the sky distribution for all black holes above $3\times 10^{7} \, M_{\odot}$, Fig.~\ref{SkyPlot2micronBLSelectionColorCodedForMassBiggerThan310To7}, then above $10^{8} \, M_{\odot}$, Fig.~\ref{SkyPlot2micronBLSelectionColorCodedForMassBiggerThan10To8}, and above $3\times 10^{8} \, M_{\odot}$, Fig.~\ref{SkyPlot2micronBLSelectionColorCodedForMassBiggerThan310To8}, also color coded in redshift.


The distribution on the sky between the cuts in black hole mass above $3\times 10^{7} \, M_{\odot}$ and above $10^{8} \, M_{\odot}$ is rather similar, with no major filamentary structure appearing or vanishing.
In contrast with the previous plot, the distribution above $3\times10^{8} M_{\odot}$ looks quite smooth on the sky with
all the major structure gone lest for some clumps of galaxies. One interesting part is the fact that from all the major filamentary structures, only the semicircular one is still present, not so evident, but easily visible. There are few galaxies that have a black hole with mass above $10^{9} M_{\odot}$ and with mass above $3\times10^{9} M_{\odot}$ only two, CGCG 522-082 with $4.5\times10^{9} M_{\odot}$ and ESO 432-IG 006 with $3.3\times10^{9} M_{\odot}$.

The distribution of the black holes and that of the super-massive ones can lead to interesting results concerning the early seeding of the universe with black holes, and their initial growth, also their distribution may relate to the nature of dark matter \citep{2005A&A...436..805M,2006A&A...458L...9M,2006PhRvL..96i1301B,2007ApJ...654..290S}.

\section{Conclusion}

The sample constructed here will allow a proper statistical treatment
 of the possible association between ultra high energy cosmic rays and the
 activity of black holes being an all sky complete collection of galaxies.

 This kind of correlations are also
 useful in testing physical theoretical models that are responsible for the
 birth and evolution of massive and super-massive black holes.

 Also this catalog of massive black hole can be use in statistical interpretation
 of the distribution of black holes with various applications from indicating the nature
 of dark matter to tracing the activity of nearby classes of galaxies.

\acknowledgement{Work with PLB was supported by contract AUGER 05 CU 5PD 1/2 via DESY/BMB and by VIHKOS via FZ Karlsruhe; by Erasmus/Sokrates EU-contracts with the universities in Bucharest, Cluj-Napoca, Budapest, Szeged, Cracow, and Ljubljana; by the DFG, the DAAD and the Humboldt Foundation; also by research foundations in Korea, China, Australia, India and Brazil. L.I.C. was supported by a research program from the LAPLAS 3 and CNCSIS Contract 539/2009.
 LIC wishes to express his thanks to the MPIfR for support when finishing this project, also discussions with A. Caramete are acknowledged. This research has made use of the NASA/IPAC Extragalactic Database (NED) which is operated by the Jet Propulsion Laboratory, California Institute of Technology, under contract with the National Aeronautics and Space Administration. This research also made use of the ViZier system at the Centre de Donne{\'e}s astronomiques de Strasbourg (CDS)  \citep{2000A&AS..143...23O}. We also acknowledge the usage of the HyperLeda database (http://leda.univ-lyon1.fr).}

\bibliographystyle{aa} 
\bibliography{references} 

\begin{table}[htpb]
\begin{center}
\footnotesize
\begin{tabular}{|c|c|c|c|c|c|c|c|c|}
\hline
Name&l&b&z&Morphological&Distance&Estimated M$_{BH}$&B-V&FIR/Radio\\
&Deg&Deg&&type&Mpc&$10^{8}M_{\odot}$&&ratio\\
\hline
\hline

NGC 5332&0.16&72.678&0.02241&S0-&92.0&3.75$\pm$1.88&0.94&-\\
NGC 6500&43.763&20.233&0.01001&SAab LINER&41.1&1.41$\pm$0.70&-&3.6$\pm$0.1\\
NGC 6849&0.329&-30.818&0.02014&SB0-&82.7&4.49$\pm$2.24&0.8&-\\
NGC 5311&83.759&72.474&0.009&S0/a&36.9&1.04$\pm$0.52&-&9.4$\pm$0.9\\
NGC 5845&0.338&48.904&0.00483&E&24.1&0.50$\pm$0.25&0.97&$<$0.17\\
NGC 5850&0.516&48.636&0.00852&SB(r)b&35.0&2.68$\pm$1.34&0.72&$<$0.819\\
NGC 7469&83.099&-45.467&0.01631&(R')SAB(rs)a Sy1.2&67.0&4.96$\pm$2.48&0.55&384.9$\pm$57.1\\
NGC 5846&0.426&48.797&0.00571&E0-1;LINER HII&29.1&5.37$\pm$2.69&0.96&$>$181.8\\
ESO 338- G 009&0.445&-23.401&0.01857&Sa-b&76.2&1.16$\pm$0.58&-&-\\
NGC 5838&0.729&49.319&0.00453&SA0-LINER&18.6&1.23$\pm$0.61&0.94&$<$0.73\\
MESSIER 094&123.363&76.007&0.001027&(R)SA(r)ab;Sy2 LINER&4.6&0.752$\pm$0.37&0.72&601.1$\pm$75.06\\
UGC 00542&123.457&-33.599&0.015044&Sb&61.7&1.32$\pm$0.66&-&$<$0.22\\
NGC 4648&123.818&42.691&0.004717&E3&19.3&0.413$\pm$0.20&0.89&-\\
UGC 00555&123.823&-33.999&0.022699&S0/a&93.2&2.67$\pm$1.34&-&-\\
UGC 01039&123.896&22.220&0.017616&Sab&72.3&1.41$\pm$0.70&-&$<$0.43\\
UGC 00567&123.897&-31.115&0.020291&S0&83.3&1.69$\pm$0.84&-&-\\
UGC 00670&123.913&12.758&0.015948&SBb?&65.4&2.33$\pm$1.17&-&$<$0.73\\
NGC 0317A&124.118&-19.057&0.017656&S0&72.5&2.09$\pm$1.04&-&$<$6.45\\
UGC 00600&124.118&-14.194&0.022726&SAB(s)b&93.3&2.06$\pm$1.03&-&$<$0.84\\
NGC 4589&124.235&42.901&0.006605&E2 LINER&20.4&1.25$\pm$0.62&0.93&7.1$\pm$0.42\\

\hline
\end{tabular}
\end{center}
\caption{Example of the data included in the massive black hole catalog, these 20 galaxies illustrate some of the different parameters that are available. We mention here that we used compiled distances as noted before where available and for all distances similar or less than the Virgo cluster the distances are corrected.}
\label{table10SMBH}
\end{table}


\begin{figure}[htpb]
\centering
\includegraphics[viewport=0cm 0cm 29cm 11cm,clip,scale=0.8]{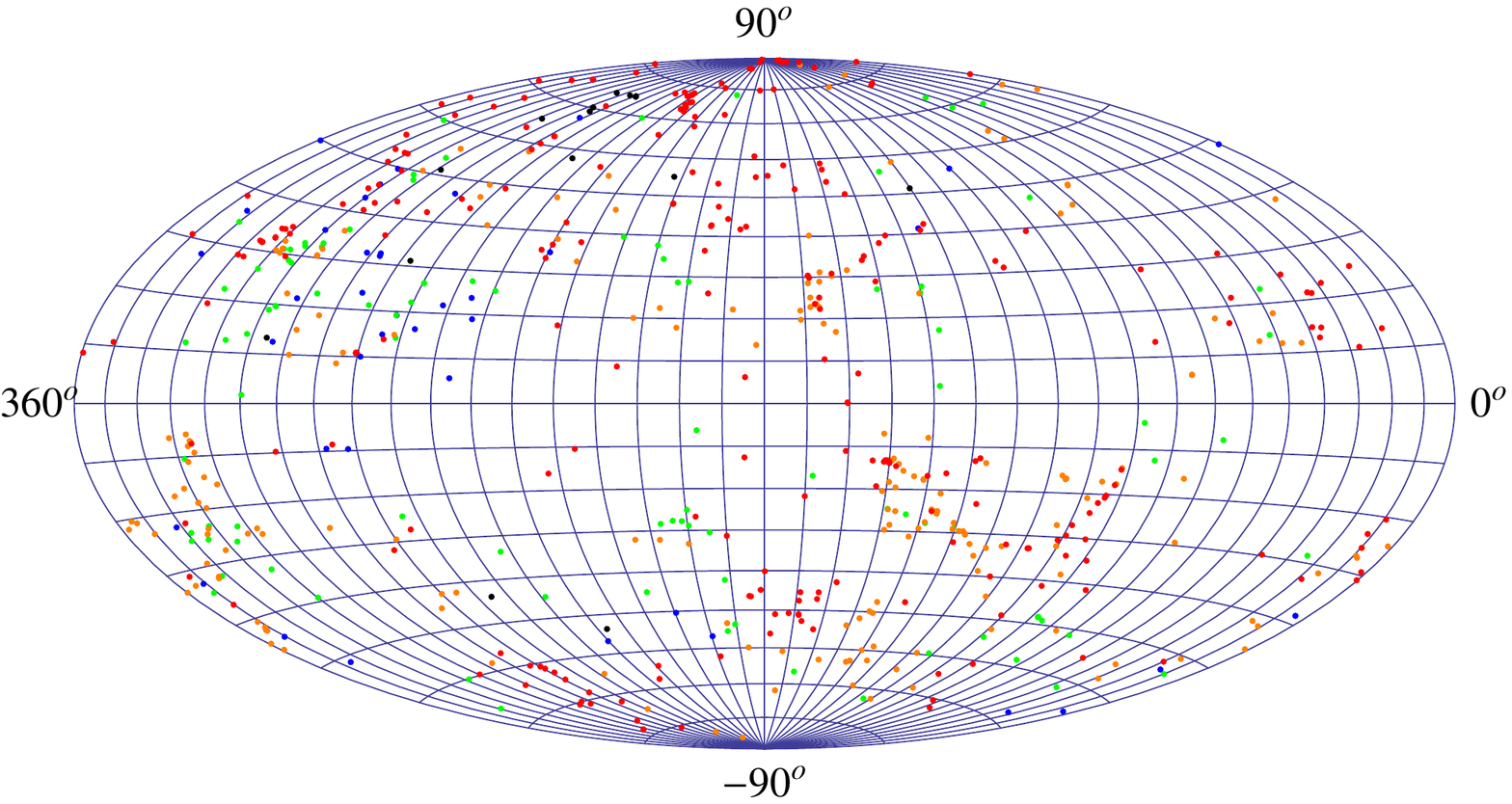}
\caption{Aitoff projection in galactic coordinates of a sample of 626 objects from the black hole catalog with masses bigger than $3\times 10^{8}\, M_{\odot}$ . The color code is Black, Blue, Green,
 Orange, Red corresponding to redshifts between 0, 0.005, 0.01, 0.015, 0.02, 0.025. }
\label{SkyPlot2micronBLSelectionColorCodedForMassBiggerThan310To}
\end{figure}

\end{document}